\documentclass[useAMS]{mn2e}
\usepackage{lscape}
\usepackage{rotating}

\def\mv{\hbox{M$_{\rm V}$}}
\def\lsun{\hbox{L$_\odot$}}
\def\msun{\hbox{M$_\odot$}}
\def\mstar{\hbox{$M_\star$}}
\def\relation{\hbox{SFR vs. M$_{V}^{\rm brightest}$}}
\def\mvten{\hbox{M$_{\rm V}^{10{\rm Myr}}$}}
\def\lir{\hbox{L$_{\rm IR}$}}
\def\msunyr{\hbox{M$_\odot$yr$^{-1}$}}

\def\cm3{\hbox{cm$^{-3}$}}

\input psfig.sty
\input epsf.sty
\usepackage{graphicx}
\usepackage{epsfig}
\title[SFR and cluster formation]
{On the star-formation rate - brightest cluster relation:  estimating the peak SFR in post-merger galaxies}
\author[N. Bastian] {N. Bastian$^{1}$ \\
$^1$ Institute of Astronomy, University of Cambridge, Madingley Road, Cambridge, CB3 0HA, UK\\
}
\date{Accepted. Received; in original form}
\pagerange{\pageref{firstpage}--\pageref{lastpage}}
\pubyear{2005}
\begin{document}
\maketitle
\label{firstpage}
\begin{abstract}
We further the recent discussion on the relation between the star-formation rate (SFR) of a galaxy and the luminosity of its brightest star-cluster (SFR vs. M$_{V}^{\rm brightest}$).  We first show that the observed trend between SFR vs. M$_{V}^{\rm brightest}$ is due to  the brightest cluster in a galaxy being preferentially young ($\leq 15$~Myr - for a constant SFR) and hence a good tracer of the current SFR, although we give notable exceptions to this rule.  Archival HST imaging of high-SFR galaxies, as well as additional galaxies/clusters from the literature, are used to further confirm the observed trend.   Using a series of Monte Carlo simulations we show that a pure power-law mass function with index, $\alpha=2$, is ruled out by the current data.  Instead we find that a Schechter function (i.e. a power-law with an exponential truncation at the high mass end) provides an excellent fit to the data.  Additionally, these simulations show that bound cluster formation (in \msun/yr) represents only $\sim8\pm3$\% of the total star-formation within a galaxy, independent of the star-formation rate.  From this we conclude that there is only a single mode of cluster formation which operates over at least six orders of magnitude in the SFR.  We provide a simple model of star/cluster formation feedback within dwarf galaxies (and star-forming complexes within spirals) which highlights the strong impact that a massive cluster can have on its surroundings.

 Using this relation, we can extrapolate backwards in time in order to estimate the peak SFR of major merger galaxies, such as NGC~7252, NGC~1316, and NGC~3610.  The derived SFRs for these galaxies are between a few hundred and a few thousand solar masses per year.  The inferred far infrared luminosity of the galaxies, from the extrapolated SFR, places them well within the range of Ultra-luminous galaxies (ULIRGs) and for NGC~7252 within the Hyper-luminous infrared galaxy regime.  Thus, we provide evidence that these post merger galaxies passed through a ULIRG/HLIRG phase and are now evolving passively.  Using the current and extrapolated past SFR of NGC~34, we infer that the ULIRG phase of this galaxy has lasted for at least 150~Myr.
\end{abstract}

\section{Introduction}

Young massive star clusters, which often surpass the globular clusters in the Galaxy in terms of brightness, mass, and density, are seen to result from intense episodes of star-formation in galaxies.
However, star clusters are also found in relatively quiescent, low star-formation rate (SFR) galaxies, albeit at much lower masses (e.g. Larsen \& Richtler~1999, 2000).  This difference in the types (mass) of clusters produced in various galactic environments has been suggested to be caused by size-of-sample effects, in which galaxies with high SFRs form proportionally more clusters, and hence are able to sample the cluster mass function out to higher masses (Larsen~2002).

This effect has been quantitatively observed through the use of the \relation~relation (Larsen~2002), where M$_{V}^{\rm brightest}$ is the brightest cluster in V-band absolute magnitude, in the sense that the most luminous clusters in galaxies with high SFRs are brighter.  This trend, along with the similar log N vs. $M_{V}^{\rm brightest}$ relation (where N is the number of clusters brighter than a certain magnitude limit; Whitmore~2003), have been used to argue for a universality of cluster formation, i.e. stochastic sampling from a universal underlying mass function.  

Size-of-sample effects, together with cluster population synthesis models (e.g.~Gieles et al.~2005) have become a common means to investigate the properties of clusters and cluster systems.  For example, Hunter et al.~(2003) used the relation of the most massive cluster per logarithmic age bin in the LMC and SMC in order to estimate the exponent of the cluster initial mass function ($\alpha$).  This proceedure was recently revisited by Gieles \& Bastian~(2008) who used the same relation to rule out mass independent, long duration ($>10$~Myr) cluster disruption models.  Gieles et al.~(2006a) used the log N vs. $M_{V}^{\rm brightest}$ relation to constrain $\alpha$, and found a value of $\sim2.4$, which is similar ($2.3$) to that derived by Weidner, Kroupa, \& Larsen~(2004) using the \relation~ relation.  This is significantly steeper than that derived from direct measurements of the mass/luminosity function of galaxies, namely $\alpha=2.0\pm0.1$ (e.g. de Grijs et al.~2003).  This discrepancy will be addressed in \S~\ref{sec:discussion}.

Wilson et al.~(2006) have tested whether the above relations still hold in the extreme environment of galaxy merger starbursts.  They studied the Ultra-Luminous Infrared Galaxy (ULIRG), Arp~220, and found that despite its high SFR ($\sim240$~\msun/yr), being an order of magnitude higher than any of the galaxies in the previous samples, falls nicely on the extrapolated fit to the more quiescent star-forming galaxies.

Weidner et al. (2004) used the \relation~relation to constrain cluster formation scenarios, namely the timescale over which clusters form, which they estimate to be on the order of a crossing time.  They further suggest that a cluster population formation epoch (i.e. the timescale where a statistically full population of clusters is formed) is on the order of 10~Myr.  However, their analysis was based on the assumption that within a "cluster population formation epoch" the brightest cluster of a galaxy is also the most massive, hence that the \relation~ trend is simply reflecting a relation between the SFR of a galaxy and the most massive cluster within it.  Observationally, it appears that this assumption is not valid, as the brightest cluster in a galaxy tends to be young, and more massive, older clusters may appear less luminous due to stellar evolution (Gieles et al.~2006a). 

In this work our goals are threefold.  The first is to test the claim by Weidner et al.~(2004) that the brightest cluster within a galaxy is also the most massive.  This naturally leads to a discussion as to why the observed \relation~relation holds.  The second is to investigate the implications of the observed relation, paying particular attention to the cluster initial mass function, and the implied connection between the cluster and star formation rates within a galaxy.  Thirdly, using the observed trend, combined with a correction for stellar evolutionary fading, to estimate the SFR in a sample of post-starburst merger galaxies.  This, in turn, allows us to place limits on the duration of the starburst phase of ULIRGS as well as trace their subsequent evolution.

In \S~\ref{sec:data} we present archival observations of two ongoing galaxy mergers and a collection of data taken from the recent literature.  \S~\ref{sec:why} presents a series of Monte Carlo simulations of cluster populations in order to investigate why the observed \relation~ relation holds.  In \S~\ref{sec:discussion} we investigate the implications for the underlying cluster initial mass function,  the relation between star and cluster formation, and use the observed relation to derive the peak SFR of post-starburst galaxies.  Our conclusions are presented in \S~\ref{sec:conclusions}.

\section{Observations and Data from the Literature}\label{sec:data}

\subsection{NGC~2623}

NGC~2623 is a luminous infrared galaxy which shows clear evidence of an ongoing merger, namely two long tidal tails and a large amount of ongoing star-formation.  It was observed with the Advanced Camera for Surveys (ACS) Wide-Field Camera (WFC) onboard HST on June 2nd, 2004 (F555W; Prop. ID 9735) and November 11th, 2005 (F435W, F814W; Prop. ID 10592).  We obtained the reduced and calibrated drizzled images through the ESO/HST archive.  We adopt a distance to NGC~2623 of 77.1~Mpc (assuming $H_{0}=72$~km/s/Mpc).

Aperture photometry was carried out (using a 10 pixel aperture and a background annulus from 12 to 14 pixels) on the brightest source in the F555W image and zeropoints from the ACS website were applied.  The brightest V-band cluster has B (F435W), V (F555W), and I (F814W) apparent magnitudes of  20.7, 20.3, and 19.6, respectively (vegamag system).  A correction for Galactic extinction was then applied (A$_{V} \sim0.1$ - Schlegel et al.~1998).  In order to estimate the extinction of this cluster we employed a $B-V$ vs. $V-I$ colour-colour plot.  Adopting the Galactic extinction law of Savage \& Mathis~(1979) we found an extinction, A$_{V}$, of 0.3~mag was necessary in order to bring the cluster colours into agreement with the SSP models.  Applying this, along with the adopted distance modulus, results in an absolute V-band magnitude of -14.5 for this cluster.




\subsection{NGC~3256}

NGC~3256 is part of the Toomre sequence of merging galaxies and has the highest current SFR and X-ray luminosity of any galaxy in the sequence.  It's known to harbor an extensive cluster population which has been studied photometrically (Zepf et al.~1999) as well as spectroscopically (Trancho et al.~2007a,b).  We adopt a distance of 37~Mpc (Zepf et al. 1999) and a Galactic foreground extinction of A$_{V}=0.4$~mag (Schegel et al.~1998).  NGC~3256 was observed with the ACS-WFC onboard HST on November 18th, 2003 (F555W - Prop. ID 9735) and November 6th, 2005 (F435W, F814W - Prop. ID 10592).  We obtained the images in the same manner as for NGC~2623.

We performed aperture photometry of the brightest cluster visable in the F555W image using the same techniques as above.  We find that the brightest cluster has  B (F435W), V (F555W), and I (F814W) magnitudes of 16.9, 17.0 and 16.36, respectively.  We have not applied any correction for intrinsic extinction.  Applying the Galactic extinction and distance modulus we find M$_{V} = -15.7$ for this cluster.



\subsection{Additional galaxies from the literature}

In Table~\ref{table:literature} we list the $M_{\rm V}$ of the brightest cluster and the estimated SFR for a sample of galaxies taken from the literature.  We have focussed our study on moderate to high SFR galaxies ($>1\msunyr$) in order to strengthen the observed trend for extrapolation to higher cluster luminosities and SFRs.  References for the brightest cluster and SFR are given, where S03 refers to a SFR estimated from the infrared luminosity (taken from Sanders et al.~2003) and SFR/L$_{\rm IR}$ relation of Kennicutt~(1998).  In the case of IRAS~19115-2124 (Vaisanen et al.~2008), we used the brightest I-band cluster ($M_{\rm I}=-17.5$; Vaisanen priv. comm.) and applied a V--I colour of 0.7 (typical of young clusters), in order to estimate the V-band magnitude.

\begin{figure}
\includegraphics[width=8.5cm]{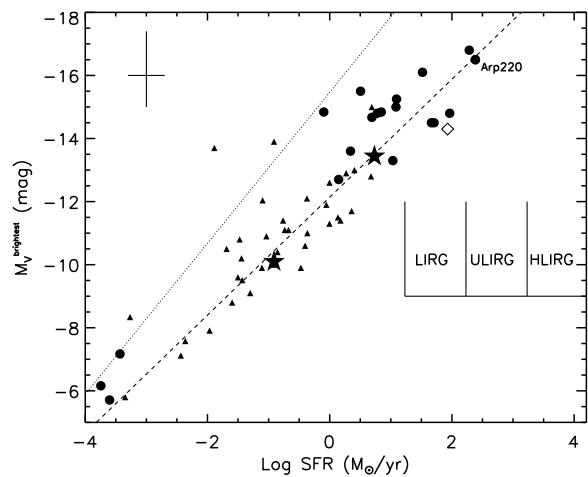}
\caption{The observed relation between the SFR of a galaxy and the V-band luminosity of the brightest cluster within the galaxy.  Galaxies taken from Larsen~(2002) are shown as filled triangles.  Solid cirlces represent galaxies from Table~1.  Filled stars represent the special cases of NGC~1569 (lower-left) and NGC~7252 (upper right), where instead of the brightest cluster of the full population, we have chosen the brightest young cluster ($<$~10~Myr).   The diamond represents the third brightest cluster in NGC~34 as the first two have ages of $\ge150$~Myr.  The best fit to the data from Weidner et al.~(2004) is shown as a dashed line.  Regions occupied by (Ultra/Hyper) Luminous Infrared Galaxies are also shown, assuming the relation between infrared luminosity and star-formation rate of Kennicutt~(1998). The dotted line shows the expected relation for a pure-power law ($\alpha=2$) case if all stars formed in bound clusters, see \S~\ref{sec:discussion} for details.  Expected error bars from stochastic sampling are shown in the upper left of the panel, along with assumed errors in the SFR.}
\label{fig:relation}
\end{figure}

The galaxy with the highest SFR galaxy in the sample is Arp~220, which was studied in detail by Wilson et al.~(2006).  They showed that this cluster fits the observed \relation~relation quite well, even though the galaxy has a SFR that is an order of magnitude higher than any other galaxies previously used in the relation.  Here we fill in that gap and show that it is indeed a continuous relation (see Fig.~\ref{fig:relation}).

In addition, we include the low-luminosity H{\sc ii} regions in the extreme outskirts of NGC~1533 which have recently been studied by Werk et al.~(2008).   These low SFR regions ($10^{-3.75} \msunyr$) are a welcome opportunity to test the low SFR regime and also test at what physical scale (i.e. galactic, H{\sc ii} region, etc) does the relation break down.  The points lie at the lower-left of Fig.~\ref{fig:relation} and can be seen to follow the extrapolated relation reasonably well.

One additional caveat is that all points in Fig.~\ref{fig:relation} are in fact lower limits in the y-direction.  The reason for this is that most of the studies used in the construction of Fig.~\ref{fig:relation} were based  on optical studies, and hence possibly effected by extinction.  Thus, it is impossible to rule out the possibility that a brighter cluster in the V-band was missed due to extinction effects.  However, the tight observed correlation suggests that this does not significantly bias the results.

\begin{table*}
\begin{centering}
{\scriptsize
\parbox[b]{12.5cm}{
\centering
\caption[]{Galaxies and clusters taken from the literature or derived in the current study.}
\begin{tabular}{c c c c}
\hline
\noalign{\smallskip}
Galaxy    & M$_V^{\rm brightest}$ & SFR &  References \\ 
                  & (mag) &   (\msunyr)  \\ 
\hline
NGC~7252$^a$ & -13.4 & 5.4 & Miller et al. 1997, Kneirman et al. 2003 \\\
ESO0338-IG04 & -15.5 & 3.2 & Ostlin et al.~1998,2007, Schmitt et al. 2006 \\
NGC~6240$^b$ & -16.4 & 140 & Pasquali et al.~2003, S03 \\
NGC~2207 & -13.6 & 2.2 & Elmegreen et al.~2001 \\
NGC~1275 & -15.3 & 12.4 & Holtzman et al.~1992, S03\\
M82 (A1) & -14.8 & 7 & Smith et al.~2006, S03\\
NGC~7673$^c$ & -14.7 &  4.9 & Homeier et al.~2002, S03\\
NGC~6745 & -15.0 & 12.2 & de Grijs et al.~2003, S03 \\
NGC~1140 & -14.8 & 0.8 & Moll et al.~2007 \\
NGC~3597 & -13.3 & 10.8 & Carlson et al.~1999, S03 \\
IRAS~19115-2124$^d$ & -16.8 & 192 & Vaisanen et al.~2008
\\
\hline
\\
Clusters in individual star-forming regions \\
NGC~1533~Assn~1  & -7.17 & 0.00037 & Werk et al.~2008, Ryan-Weber et al.~2004\\
NGC~1533~Assn~2  & -5.71 & 0.00025 & "\\
NGC~1533~Assn~5  & -6.16 &0.00018 & "\\

\\
\hline
\\
Clusters presented in this work \\
NGC~2623 & -14.5 & 51. & S03 \\
NGC~3256 & -15.7 & 46 & S03 \\
\noalign{\smallskip}
\noalign{\smallskip}
\hline
\end{tabular}
\begin{list}{}{}
\item[$^{\mathrm{a}}$] The brightest cluster in the central star-forming region of the galaxy.  See text for details.
\item[$^{\mathrm{b}}$]  We have used the brightest cluster in their sample and corrected for E($B-V$)=0.4 mag which is the average extinction for the main body of the galaxy.  We have also excluded the nuclear star-forming region (three knots) as they are likely to have extended star-formation histories (Pasquali et al.~2004).
\item[$^{\mathrm{c}}$] Corrected for an extinction ($A_{\rm V}$) of 0.2~mag. 
 \item[$^{\mathrm{d}}$] See text for details on the derivation of $M_{\rm V}^{\rm brightest}$.
\end{list}
\label{table:literature}
}
}
\end{centering}
\end{table*}

\section{The Relation Between the SFR and the Brightest Cluster}
\label{sec:why}

The relation between the SFR of a galaxy and the brightest cluster within it is empirically based, and was originally given a statistical (i.e. size-of-sample) explanation (Larsen 2002).  However, Weidner et al. (2004) suggested that the underlying cause was that a full, statistically complete cluster population was formed every $\sim10$~Myr, and that this rapid formation timescale was at the basis of the observed SFR/cluster relation.   For this the authors assumed that the most massive cluster in a population is normally also the brightest.

In order to test this assertion we performed a simple series of Monte Carlo simulations of cluster populations.  We create 100 realisations of a cluster population in a galaxy.  The clusters are drawn stochastically from a mass function (a power-law with an index of $-2$) with a lower limit of 100~\msun~ and an upper limit $10^{12}$~\msun (effectively no upper limit), and originally we create 5000 clusters. The clusters are then assigned ages randomly between 0 and 100 Myr (in order to simulate a constant cluster formation rate).  We then calculate the absolute magnitude of each cluster, using the cluster's age and mass, from simple stellar population models (Bruzual \& Charlot~2003; solar metallicity, Salpter stellar IMF).  Once this is carried out for a single realisation, we search the population for the brightest cluster and the most massive (which are not necessarily the same cluster) along with the corresponding clusters' ages.  The average cluster formation rate (CFR) is found by dividing the total mass formed in clusters by the duration of the experiment (i.e. 100~Myr).

\begin{figure}
\includegraphics[width=8.5cm]{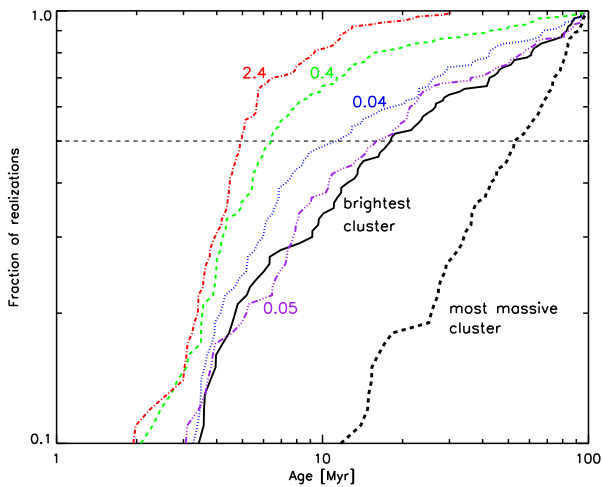}
\caption{The results of Monte Carlo experiments of a cluster population with a power-law mass function with index -2 (solid line and thick dashed line) or a Schechter mass function (other lines).  The curves represent the cumulative distribution of the fraction of realizations where the brightest cluster is younger than a given age, or in the case of the right solid line, the most massive cluster.  The 50-percentile is marked with a horizontal dashed line.  The numbers next to the curves show the cluster-formation rate (in \msun/yr) assumed in each simulation (the power-law results are independent of the assumed CFR).  The results shown are lower limits since we have not included any mass loss in the simulations, only stellar evolution.  Additionally, we show the cumulative distribution of the age where the most massive cluster appears as a thick dashed line.  This is clearly weighted towards older ages, hence the brightest cluster is often not the most massive.}
\label{fig:mc}
\end{figure}

Fig.~\ref{fig:mc} shows the cumulative distribution of the fraction of realizations whose brightest cluster is younger than a given age (solid black line).  We also show the age distribution of the most massive cluster in each sample.  The 50-percentile mark is also shown, which we see for this simulation happens at $\sim15$~Myr for the brightest cluster (i.e. in 50\% of the realizations the brightest cluster had an age of 15~Myr or less) and $\sim55$~Myr for the most massive cluster.  Gieles~(2008) has shown that if these simulations were to be carried out over a much longer timescale (e.g. a few Gyr) the brightest cluster would usually be old, as the most massive cluster increases faster (due to the size-of-sample effect) than the fading of clusters past a few 10s of Myr.  However, in reality a cluster is expected to lose a large fraction of its mass over time due to tidal and internal effects, which is why we limit our simulation to 100~Myr.   We note that if we use the {\rm STARBURST99} models (Leitherer et al.~1999) the results remain unchanged.

From this we conclude that for a power-law mass distribution (with an index of $-2$) at least half of the time the brightest cluster will be young and not necessarily the most massive.  Note that we have not included the effects of infant-weight loss (Goodwin \& Bastian 2006) nor cluster mass loss due to tidal effects (e.g. Lamers et al.~2005, Gieles et al.~2006c), both of which would make older clusters less massive (on average) and hence decrease their luminosity faster than would be predicted based solely on stellar evolution.  Thus, the shown distribution is in fact a lower-limit.  We also conclude from Fig.~\ref{fig:mc} that the most massive cluster in a galaxy is likely to be old, hence the brightest cluster is often not the most massive cluster.

These results differ from those of Weidner et al~(2004) due to the duration of the experiment.  If we would use a duration of 500~Myr, instead of 100~Myr, then we reproduce their results, namely that in $\sim95$\% of the cases the most massive cluster is also the brightest.  This is because over long time spans, the mass of the most massive cluster increases due to size-of-sample effects faster than stellar evolutionary fading reduces its brightness (Gieles~2008).  Observationally, however, the most massive cluster in a galaxy does not continue to increase at the rate expected for the size-of-sample effect, but grows more slowly after $~100$~Myr (Gieles \& Bastian~2008), presumably due to disruption effects.

Additionally, we have tested the expected distribution (fraction of cases where the brightest cluster is young) for an underlying Schechter mass function.  This type of distribution behaves as a power-law in the low-mass regime (which we assume to have an index of $-2$), but decreases more rapidly at high-masses a certain characteristic mass, \mstar.  We carried out the same set of simulations as above, but using a Schechter mass function with \mstar$=2\times10^{6}$~\msun. This particular limit is similar to that found by Gieles et al.~(2006a,b) and used by Jord{\'a}n et al.~(2005), and will be justified further in \S~\ref{sec:cimf}.  We have run a series of simulations with different cluster-formation rates (CFR), which is necessary as the higher the CFR the more clusters are sampled near/above \mstar.  The results are shown in Fig.~\ref{fig:mc} where the number next to each curve is the CFR (in \msun/yr).  The dash-dot-dot-dot (purple) line represents a low CFR (0.05 \msun/yr) which, as expected, is very similar to the pure power-law case, since it does not sample near \mstar.  Not surprisingly, as the CFR increases and one samples closer to, and above, \mstar\, the maximum cluster mass does not increase rapidly.  Hence the population will be dominated by youth, rather than mass.  For high CFR ($>1$~\msun/yr)  the chance of finding an older cluster which is the brightest in a galaxy essentially drops to zero.

{\it Thus we are left to conclude that the tight relation between the observed SFR of a galaxy and the brightest cluster in the population is due to the fact that the youngest clusters are generally the brightest, and hence accurately reflect the current (recent) SFR.}

\subsection{Exceptions that prove the rule - the example of NGC~7252}
\label{sec:example}

The above experiments assumed that the CFR of a galaxy was approximately constant during the duration of the simulations (100~Myr).  However, this assumption clearly does not hold for starbursts, in which a galaxy can undergo a significant enhancement of its SFR for extended periods.  To demonstrate this effect we use the merger remnant NGC~7252 as an example.  This galaxy is likely to have been produced by an equal mass gas-rich spiral/spiral galaxy merger  (Hibbard \& Mihos 1995) approximately 500-1000~Myr ago (Schweizer \& Seitzer~1998).   NGC~7252 is an outlying point in the \relation~ relation, being significantly above the fit to the other galaxies ($>4\sigma$\footnote{Where $\sigma$ is calculated from the difference between the observed $M_{\rm V}^{\rm brightest}$ and the best fit relation and error by Weidner et al.~(2004).}, i.e. its brightest cluster is too bright for its current SFR), and as such, was not included in the fit by Weidner et al.~(2004).

During this merger, many extremely massive star clusters were formed, including a handful of clusters with masses exceeding $10^7$ \msun\ (Schweizer \& Seitzer 1998, Maraston et al.~2004, Bastian et al.~2006b).  After this strong starburst, the SFR has presumably been decreasing to its present rate of 5.4~\msunyr (Kneirman et al.~2003), which is localized mainly to a relatively small inner star-forming disk (Schweizer~1982).  The current SFR is $\sim3$ orders of magnitude less than the implied SFR (see \S~\ref{sec:sfr}), hence the currently forming clusters are expected to be much less massive than those formed during the main starburst event.  Thus, even though these clusters are much younger than the clusters formed in the burst, they are fainter, given the extremely high masses of the clusters formed in the merger.

If one were to use the brightest {\it young} cluster in NGC~7252 (which is found in the star-forming disk), which has $M_V = -13.4$ (Miller et al.~1997) then this galaxy sits well on the SFR-brightest cluster diagram, shown in Fig.~\ref{fig:relation} (the filled star towards the upper-right of the figure).

The conclusion is then that the {\it current} SFR of a galaxy can be inferred from the absolute magnitude of the brightest {\it young} cluster of the population.  In \S~\ref{sec:discussion} we will expand this idea and infer the peak SFR of a galaxy merger by the (evolutionarily corrected) luminosity of its brightest cluster.

\subsection{A model for the evolution of dwarf galaxy starbursts in the \relation~ plane}
\label{sec:dwarfs}

It has been noted that dwarf/irregular starburst galaxies often lie significantly above the relation shown in Fig.~\ref{fig:relation} (e.g. Larsen 2002, Billett, Hunter \& Elmegreen~2002), a fact which led Weidner et al.~(2004) not to include them in their fit to the data (shown as a dashed line in Fig.~\ref{fig:relation}).  A number of theories have been put forward to explain this offset.  Billett et al.~(2002) suggest  that this offset is due to different conditions governing cluster formation in small galaxies.  Weidner et al.~(2004), on the other hand, attribute this offset to star-formation proceeding in short bursts in small galaxies (e.g. Searle \& Sargent~1972).  Here, we use the dwarf irregular (post) starburst galaxy, NGC~1569, to argue for the latter model, which we summarize below.

\begin{figure}
\includegraphics[width=8.5cm]{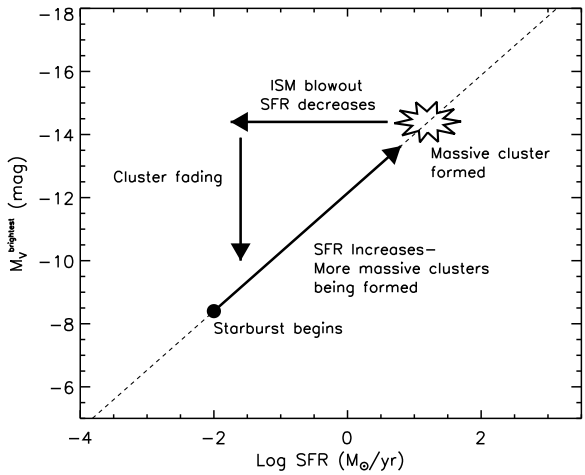}
\caption{A schematic representation of the evolution of a dwarf galaxy starburst in the SFR vs. M$_{V}^{\rm brightest}$ plane. }
\label{fig:dwarf-evo}
\end{figure}

One clear difference between spirals and smaller dwarfs is the amount of energy and the timescale required to disturb the galaxy as a whole.  In the case of small galaxies, a single burst of star-formation may be enough to completely rid the galaxy of any gas reservoir.  Studies of NGC~1569 have shown that the inner region of the galaxy hosts two massive ($>10^5$~\msun) clusters, with ages between 15-30~Myr (Anders et al.~2004, Larsen et al.~2008).  The inner region is currently largely devoid of dense star-forming gas and a powerful supergalactic wind is seen centred on the central region (e.g. Westmoquette et al.~2007) which was presumably powered by the energy input into the ISM from these two massive clusters.  Thus, as the galaxy began to form stars/clusters more vehemently, it eventually reached the point where a single (or in this case two) massive clusters formed, which caused a blowout of the ISM within the central regions of the galaxy.  The SFR then dropped due to the lack of dense gas, and henceforth only smaller clusters could be formed.  A schematic diagram of this process is shown in Fig.~\ref{fig:dwarf-evo}.  In such a model, galaxies which fall to the left of the observed trend in Fig.~\ref{fig:relation} should have older  brightest clusters than the average cluster used in the diagram, i.e. have ages larger than $10-20$~Myr, so that enough time has passed for the clusters to have had a large influence on the surrounding ISM.

Additional evidence can be found in the young stellar clusters in this galaxy.  Anders et al.~(2004) have catalogued and age-dated approximately 160 clusters in NGC~1569.  Using their dataset, and looking at only those clusters with ages less than 10~Myr, we find that the brightest young cluster in the galaxy has M$_{\rm V}=-10.31$.  It is this cluster which should be used in comparison with the current SFR.  In Fig.~\ref{fig:relation} we show the position of this cluster in the \relation diagram as the lower filled star.  This cluster appears to fit the relation splendidly, hence arguing that the cluster formation process is not intrinsically different in dwarfs and spirals.

NGC~1705, the other dwarf post-starburst galaxy which lies significantly to the left of the observed relation, also has had an explosive outflow attributed to the formation of its most massive cluster which occured $\sim12$~Myr ago (V\'azquez et al.~2004).  This galaxy as well seems to have drastically reduced its SFR through the feedback associated with forming a massive clusters.

However, we note that NGC~1140 appears to be 1.5-2$\sigma$ off the expected \relation~relation.  This galaxy has been classified as an amorphous galaxy (Hunter et al.~1994), having recently undergone a strong interaction.  However, it may have originally been a low-luminosity late type spiral (Hunter et al.~1994).   This galaxy has a current SFR of about 0.8~\msunyr and a brightest cluster of $M_{\rm V}= -14.8$ which has an age of 4--7~Myr (Moll et al.~2007).  Due to the youth of the cluster, we would expect it to accurately reflect the current SFR in the galaxy.  If this cluster/galaxy represents a statistical fluctuation or a real physical difference cannot be concluded at this time.

This conceptual model makes two predictions.  The first, is that there should be few/no galaxies which lie significantly below the observed \relation\ relation (i.e. more than random sampling would predict), only above.  The second is that galaxies which lie significantly above the relation should have brightest clusters which are older than the average age of clusters whose galaxies fit the relation.

\subsection{Intense star-forming complexes within galaxies}

This same effect, i.e. feeback from massive stars/clusters causing a drop in the SFR, is likely to happen within individual star-forming regions within spiral galaxies (e.g. Shetty \& Ostriker~2008), like those found in M51 or the Antennae galaxies (Bastian et al.~2005, 2006a).  Here the important length scale is that of the GMC.   Once the GMC begins to form clusters, the clusters, through radiative feedback, destroy the host GMC, and eventually halt star-formation within the region/complex.  Thus we would predict older complexes to lie to the left of the observed \relation~relation while young ones (i.e. currently forming) to lie on the relation, where the SFR should be estimated over the region of interest.  

\begin{table*}
\begin{centering}
{\scriptsize
\parbox[b]{18cm}{
\centering
\caption[]{The properties of galactic merger remnants.}
\begin{tabular}{c c c c c c c c}
\hline
\noalign{\smallskip}
Galaxy   & cluster ID& M$_V^{current}$ &  M$_V^{10 Myr}$ & Age (Myr) & peak SFR & peak log (\lir) &  References \\ 
                  &  &  (mag) &  (mag) & (Myr)    &    (\msunyr) & (\lsun) \\ 
\hline
NGC~7252 & W3 & -16.2 &-18.9 & 400 & 4131 & 13.4 & Miller et al.~1997\\
NGC~7252 & W30$^a$ & -14.6 & -17.3& 400 & 576 & 12.5 & Miller et al.~1997 \\
NGC~1316 & G114 &-13.0  &  -17.6 & 3000 & 853 & 12.7 & Goudfrooij et al.~2004\\
NGC~3921& S1 &-12.5 & -15.3 & 450&  50 & 11.5 & Schweizer, Seitzer, \& Brodie.~2004\\
NGC~3610 & \#1 & -11.6 & - 16.5 & 4000 & 218 & 12.1 &Whitmore et al.~2002\\
NGC~1700 & \#1 &-10.9& -15.8 & 4000 &91 & 11.7 & Whitmore et al.~1997, Brown et al. 2000\\
NGC~34 & \#1& -15.4 & -17.3 &150 &  601 & 12.5 & Schweizer \& Seitzer 2007\\
NGC~3597 & - & -13.2 & -16.4 & 200 & 57 & 11.5 & Holtzman et al.~1996\\
\hline
\end{tabular}
\begin{list}{}{}
\item[$^{\mathrm{a}}$] This is the second brightness cluster in the galaxy, given here as an example of the uncertainties in the method.
\end{list}
\label{table:mergers}
}
}
\end{centering}
\end{table*}

\section{Discussion and Implications}
\label{sec:discussion}

\subsection{The cluster initial mass function}
\label{sec:cimf}

The \relation or equivocally N$_{\rm clusters}$ vs. M$_{V}^{\rm brightest}$ relations have been used previously in order to constrain the initial mass function of clusters.   Multiple studies have reported that the observed correlation is consistent with a power-law mass function with index,  $\alpha = \sim2.3-2.4$ (Whitmore~2003;  Larsen 2002; Weidner et al.~2004; Gieles et al. 2006a).  
This, however, is inconsistent with direct measurements of the mass function which generally find $\alpha=2.0$ (e.g. Zhang \& Fall~1999; de Grijs et al.~2003, McCrady \& Graham~2007).

Gieles et al.~(2006a) have suggested that instead of a power-law mass function, the underlying mass distribution of clusters is better described by a Schechter function.  This gives the standard $\alpha=2$ form on the low mass side, but is truncated on the high mass side above some critical value \mstar.  This type of function is physically motivated, as GMCs (i.e. the material from which clusters form) appear to have a power-law mass function which is truncated at the high mass end (e.g. Rosolowsky et al. 2007).

In order to test this assertion we develop the following model.  We generate 50 cluster populations for eleven cluster formation rates (CFR) and stochastically  sample an underlying Schechter mass function with a lower cluster mass limit of 100~\msun, a truncation mass \mstar, and an upper mass limit $M_{\rm up} >> \mstar$.  Each population is generated with $N$ clusters which are assigned ages randomly between 0 and 100 Myr.  We then calculate the absolute magnitude of each cluster, in the same way as was done in \S~\ref{sec:why}.
 Finally, we calculate the average {\it cluster formation rate} (CFR) by dividing the total mass in clusters formed in each population by the age range sampled (i.e. 100~Myr).

\begin{figure}
\includegraphics[width=8.5cm]{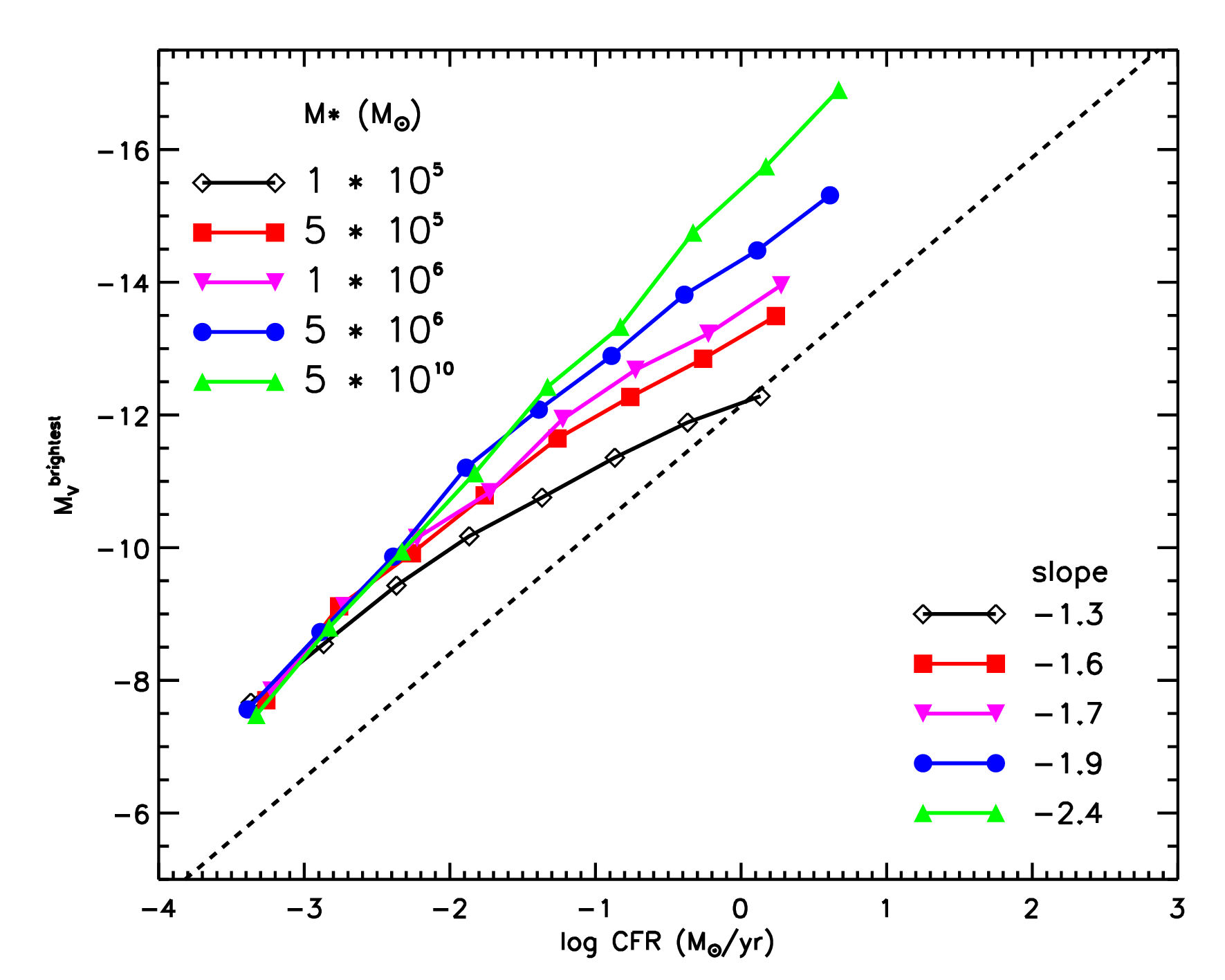}
\includegraphics[width=8.5cm]{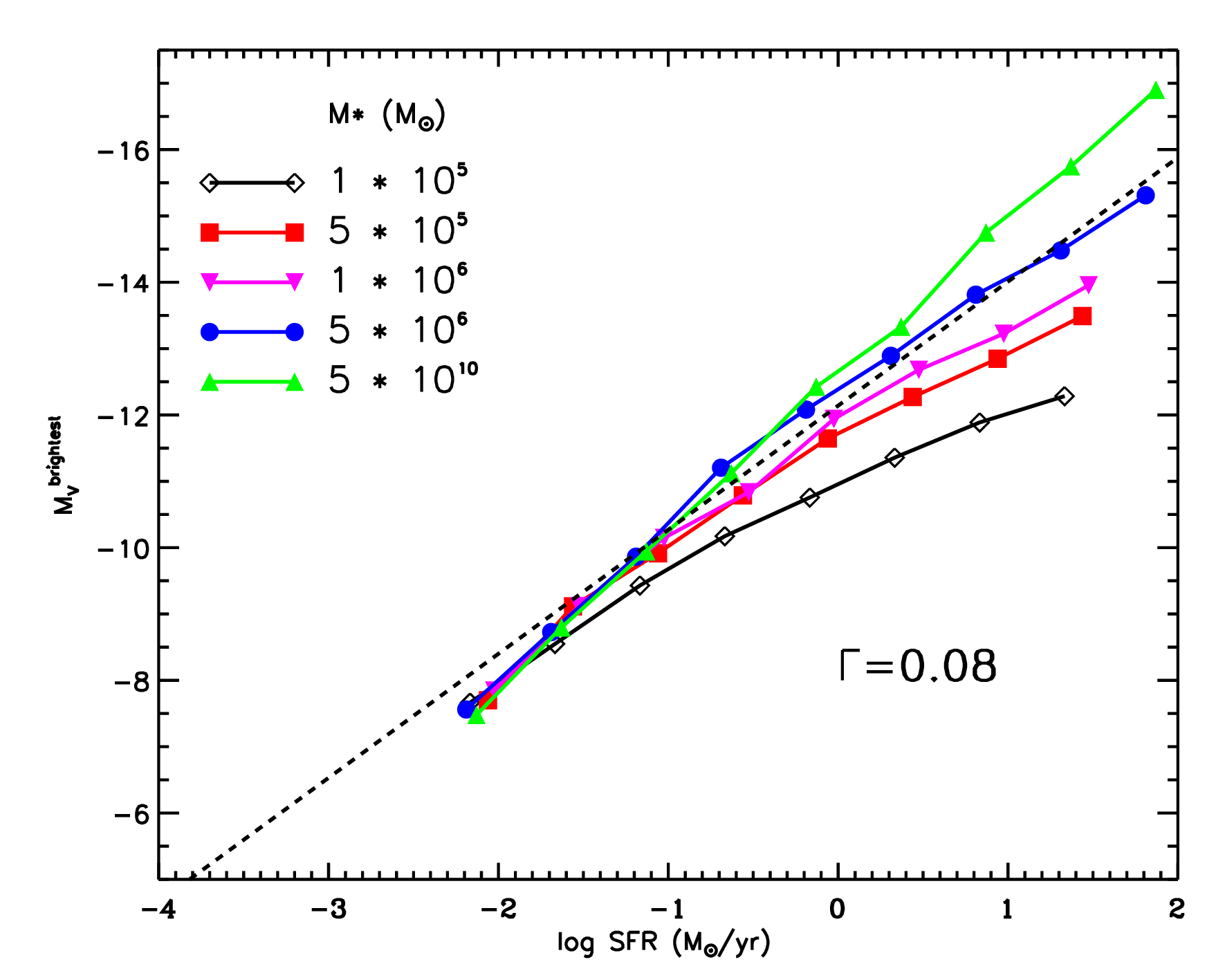}
\caption{The results of a series of Monte Carlo simulations of cluster populations, where solid lines represent the average results of 50 realizations of 11 cluster formation rates (i.e. each line represents 550 simulations).  For each simulation we assume a constant cluster formation rate and sample the mass distribution stochastically from a Schechter function with varying values of \mstar.  {\bf Top:}  The results of the simulations for the {\it cluster formation rate}  Note that the simulations are significantly to the left of the observations, implying that clusters only represent a small fraction of the total star-formation of a galaxy. {\bf Bottom:}  The same as the top panel, but now with the factor $\Gamma=0.08$ applied, i.e. converting the CFR to SFR.   The dashed lines shows the observed relation as fit by Weidner et al.~(2004), which has a slope of $-1.87$.  The filled triangles represent a case where \mstar~ is sufficiently high as to not affect the mass distribution (i.e. a representation of the pure-power law mass function case).  As can be seen, a Schechter function with \mstar$=1-5\times10^6$~\msun~ provides an excellent fit to the data, while the pure power-law case over-predicts the luminosity of the brightest cluster for a given SFR.}
\label{fig:relation-sim}
\end{figure}

The top panel of Fig.~\ref{fig:relation-sim} shows the results of the simulations for various values of \mstar, where the points and lines represent the mean in the logarithm of the SFR and magnitude of the 550 realizations for each value of \mstar.   The values of \mstar~ and the resulting slopes from linear least-square fits to the unbinned data points (i.e. in the same way that Weidner et al.~(2004) have fit the data), are also shown in the figure.  The bottom panel of Fig.~\ref{fig:relation-sim} shows the same simulations but with the CFR translated into a SFR using $\Gamma=0.08$ (the fraction of star-formation which happens in bound clusters), 
which will be discussed in detail in \S~\ref{sec:cfr-sfr}.  The dashed line in both panels is the observed relation, taken from Weidner et al.~(2004), which has a slope of $-1.87$.  
 As can be seen, low values of \mstar~ ($1-5 \times 10^{5}$~\msun)  do not fit the data well in the high SFR regime, and high \mstar~ values (i.e. pure power-laws with index of $-2.0$) over-predict the magnitude of the brightest cluster for a given SFR.  However, values of \mstar~ of $1-5 \times 10^{6}$~\msun~ appear to fit the observed trend quite nicely.
 
 Weidner et al.~(2004) found that the observations were best fit by M$_{\rm V}^{\rm brightest} \propto -1.87(\pm0.06) \times {\rm SFR}$.  This is $>8\sigma$ off that expected from the pure power-law, $\alpha=2$ case (seen as the green line with triangles in Fig.~\ref{fig:relation-sim}),  hence we conclude that this is ruled out by the current data.

The rather high values of \mstar~ which do fit the data ($1-5 \times 10^6$~\msun) make a direct comparison (i.e. a binned mass distribution) between a power-law and a Schechter mass function extremely difficult, as the two populations only differ by a handful of the most massive clusters.  An example of this is found in Whitmore, Chandar, \& Fall.~(2007) who found a lack of massive clusters at the high mass end of the mass distribution in the Antennae galaxies compared to a pure power-law with an index of $-2$, and they note that an index of $-2.1$ to $-2.2$ is required.  This difference was near the Poissonian limit so no strong conclusion could be reached.  

However, size-of-sample effects, like the \relation~ relation, which rely on only the brightest cluster in a sample, are quite sensitive to the difference between the two functions in the high mass regime, allowing for a clear separation.

It should be noted that it is unlikely that the value of \mstar~ is universal and that it probably depends on the ambient conditions of the host galaxy, i.e. to what mass can the galaxy form the molecular concentrations required to form star clusters.  For example, the truncation in the GMC mass function in the inner regions of M33 happens at $\sim8\times10^5$~\msun~ (Rosolowsky et al.~2007), which would lead to a cluster mass truncation below this value.  In contrast, major galaxy mergers such as NGC~7252 and Arp~220 may be able to form extremely dense and high mass gas concentrations.  A Schechter function with \mstar~ of $1-5\times10^6$~\msun~ can be ruled out in the case of NGC~7252 due to the large number of clusters with masses well in excess of this value (Bastian et al. 2006b).   Additionally, a Schechter function results in a slightly curved \relation\ relation where at low SFR (when a population does not sample near \mstar) the slope is the same as for the power-law exponent in the Schechter function.  If \mstar, however, weakly correlates with the SFR (as might be expected if \mstar\ is dependent on galaxy type) then the slope would be preserved beyond the plotted bounds in Fig.~\ref{fig:relation-sim}.

\subsection{The relation between the cluster and star formation rates}
\label{sec:cfr-sfr}

In order to compare our Monte Carlo cluster population simulations to the observed data, we need to apply a correction factor between the cluster-formation rate (CFR) and the star-formation rate (SFR), which we define as $\Gamma \equiv {\rm CFR/SFR}$.  This factor enters as a horizontal scaling in Fig.~\ref{fig:relation-sim} (i.e. the horizontal shift between the top and bottom panels).  In order to equate our best fit simulations (in terms of the slope of the relation, i.e. a Schechter function with \mstar$=1-5\times10^6 \msun$)  to the observations we needed to use $\Gamma = 0.08\pm0.03$, meaning that optically selected "bound" clusters represent only 8\% of the total star-formation of a galaxy.   If we instead adopt a pure-power law with index $\alpha=2.3$ then $\Gamma = 0.2$.  By "bound" we refer to clusters which have survived the transition from being embedded in molecular gas to being exposed, and hence possible to be included optically selected cluster samples (see Goodwin~2008 for a recent review of this transition period).  Additionally, the cluster must be compact enough so as to make it into cluster samples, for which it is often assumed that the cluster must then be gravitationally bound.  Note, however, that we cannot constrain the fraction of stars which are {\it formed} in dense clusters relative to the field or loose association.  We can only constrain the fraction of stars, relative to the total, which end up in bound clusters after the transition for embedded to exposed.

It is also possible to shift the other simulations onto the observations using a conversion factor which is dependent on the SFR.  However, for the pure power-law case with an index of $-2$, this implies that cluster formation becomes {\it less} efficient for high SFR.  This issue is discussed in detail in Bastian \& Gieles (in prep.).  \footnote{We note that these results are largely independent of the adopted lower cluster mass limit (for which we use 100~\msun), since the CFR is determined after each population has been constructed (total mass in clusters divided by the duration of the experiment).  Hence for a given number of clusters generated, decreasing the lower limit, decreases the CFR and decreases the expected M$_{V}^{\rm brightest}$, hence shifting the curves along their slopes.}


This value of $\Gamma$ is in good agreement with that found for the solar neighbourhood by Lada \& Lada ($\Gamma=0.04-0.07$; 2003) and Lamers \& Gieles~($\Gamma = 0.05-0.11$; 2007).  It is also in relatively good agreement with that found by Gieles \& Bastian (2008) for the SMC ($\Gamma=0.02-0.04$).  The interpretation and implications of such a low value of $\Gamma$ are also treated in detail in Bastian \& Gieles~(in prep.).
 
Interestingly, the same 'efficiency factor' was needed, for all values of the CFR, in order to fit the data. This means that the same fraction of stars which are forming are going into clusters, independent of the SFR.  Along with the smooth trend in Fig.~\ref{fig:relation}, {\it this implies that there do not exist multiple modes of cluster formation}.  Starburst galaxies do not preferentially form high-mass clusters, they are simply able to sample further out into the cluster mass function.

Since the CFR appears to make up a more or less constant fraction of the SFR (for $10^{-4} <$ SFR (\msun/yr) $< 240$), it naturally follows that the star and cluster formation histories of a galaxy should coincide.  
Exceptions to this ansatz may be due to the combined uncertainties of the cluster disruption correction and the sampling of small fields for the star-formation history which can suffer from severe variance (e.g. Maschberger et al., in prep).


\subsection{Inferring the SFR of galaxies in the past}
\label{sec:sfr}

Combining equation~2 in Weidner et al.~(2004), which is shown as a dashed line in Fig.~1, and equation~4 in Kennicutt~(1998), one can find the relation between the brightest cluster, the SFR, and the infrared luminosity of a galaxy.   However, as noted above, the brightest star cluster {\it associated with that epoch of star-formation} must be used.   By using the brightest cluster in a galaxy, and accounting for cluster fading due to stellar evolution using simple stellar population models, one can obtain an estimate of the SFR during the epoch of the formation of that cluster.  This is similar to what has been done by Maschberger \& Kroupa~(2007), however, instead of deriving the full star-formation history of a galaxy, we are interested in what was the peak SFR in that galaxy's (recent) history.

For such a study, post-galaxy mergers are an ideal place to look, as they are known to form copious amounts of massive clusters (e.g. Schweizer~1987, Ashman \& Zepf~1992).  Since major mergers of galaxies are relatively rare in the local universe, one can look at the fossil remnants which are much more prevalent, in order to infer the conditions during the merger.  Many nearby ($<100$~Mpc) merger remnants have had their star-cluster populations catalogued and here we will estimate the SFR and peak log (L$_{\rm IR}/\lsun$) in order to see the prevailing conditions at the time of the cluster formation and whether these galaxies passed through a ULIRG/HLIRG phase.

Combining the observed \relation~ relation and that between the SFR and far-infrared luminosity (Kennicutt~1998) results in the equations:
\begin{equation}
{\rm SFR (\msunyr)} = 10^{\frac{M_{\rm V}^{\rm brightest} + 12.14}{-1.87}}
\end{equation}
\begin{equation}
{\rm SFR (\msunyr)} = 4.5 \times 10^{-44} L_{\rm FIR} ~{\rm (ergs~s^{-1}) }
\end{equation}
\begin{equation}
\frac{M_{\rm V}^{\rm brightest} + 12.14}{-1.87} ={\rm log}~(4.5 \times 10^{-44} L_{\rm FIR})
\end{equation}

The results for seven post-merger galaxies are given in Table~\ref{table:mergers}.  We have used the published values of $M_{\rm V}$ and cluster age, and used the Bruzual \& Charlot~(2003) SSP models (solar metallicity and Salpeter stellar IMF) in order to estimate the absolute magnitude of the brightest cluster at an age of 10~Myr, $M_{\rm V}^{\rm 10 Myr}$ .  The estimated values of \mvten are lower limits since we have not attempted to correct for mass loss due to tidal/dynamical effects, which are expected to be relatively small since the mass-loss is inversely related to cluster mass (e.g. Baumgardt \& Makino~2003, Lamers et al.~2005).

Additionally, the peak \lir~ values quoted here are also lower limits since they do not consider any AGN component which may have been present.  It has been noted (e.g.~Veilleux, Sanders, \& Kim~1999) that the fraction of galaxies with a clear AGN component increases with increasing luminosity.  However,  Verma et al.~(2002) note that in their sample of four HLIRGs, three needed substantial fraction of the \lir~ to come from an obscured starburst (from 30--75\% of the total \lir~ luminosity).

NGC~7252 stands out among the others in its estimated star formation rate ($>4000$~\msunyr) and its peak luminosity would clearly place it in the range of Hyper-luminous infrared galaxies (HLIRGs).  Verma et al.~(2002) have decomposed the infrared spectra of two HLIRGs into their starburst and AGN components, and have estimated star-formation rates above 3000~\msunyr.  If, instead of W3, we used the second brightest cluster in the NGC~7252, W30, the implied SFR and peak \lir~ would be $\sim600$~\msunyr~ and $10^{12.5}$\lsun, respectively.  While still well within the ULIRG category, this shows the sensitivity and limitations of the method used here.  One possibility to decrease this sensitivity would be to use the say three or five brightest clusters, which would reduce the sampling scatter.

Young clusters do not form in isolation but are often, instead, found in cluster complexes (e.g. Bastian et al.~2005,2006).  Within these complexes, cluster merging may be an important mechanism in building massive clusters (e.g. Fellhauer \& Kroupa~2002; Kissler-Patig, Jordan, \& Bastian~2006).  Through merging, it is possible to raise a galaxy in the \relation~relation, as the brightness of the most luminous cluster would increase through cluster mergers, but the SFR would remain unchanged.  We do not expect this to bias our results substantially, as the \relation~relation is empirically derived, and this effect may already be included.  However, if a significant amount of merging happens after the first 15--20~Myr of a cluster's life, the current (stellar evolutionary corrected) magnitudes would be over-estimates leading to high SFR and \lir\ derivations.





\subsection{Duration of the LIRG/ULIRG phase}

One galaxy, NGC~34, stands out as unique in the sample assembled here.  It is currently a LIRG with a far-IR luminosity of 10$^{11.44} \lsun$, which translates to a SFR of 75-90 \msunyr (S03; Prouton et al.~2004).  However, it appears to have had a higher SFR in the past.  Schweizer \& Seitzer~(2007) have obtained spectroscopic ages of the two brightest clusters (\mv$=-15.36$) in the galaxy, and have derived ages of $\sim150$~Myr (although they note that cluster~2 may be $\sim4$ times older).  Correcting for evolutionary fading and applying Eqns.~$1-3$ gives an absolute V-band magnitude at 10~Myr, peak SFR, and peak \lir~ of -17.3 mag, $\sim600$~\msunyr, and $10^{12.5}~\lsun$, respectively.  Therefore, we can infer that the SFR and \lir~ have both been declining for the past 150~Myr to the current levels, and that this timescale is a lower limit to the duration of the LIRG/ULIRG phase.  If cluster~2 is indeed four times older, then the ULIRG duration for this galaxy is at least $\sim600$~Myr,  or if the SFR has not been continuous, then NGC~34 has gone through multiple ULIRG phases.

\section{Conclusions}
\label{sec:conclusions}

Using the empirically derived relation between the star-formation rate (SFR) and the luminosity of the brightest star cluster within the galaxy (Larsen 2002) we have investigated the conditions (SFR and infrared luminosity) that were present during the peak star-forming epoch of galactic mergers.    We began with a series of simple Monte Carlo simulations of cluster populations with different assumptions on the underlying cluster mass function, and showed that, independent of the model construction, the youngest clusters ($<10$~Myr) were predominantly the brightest.  This, along with size-of-sample effects, can explain the observed \relation~ relation, as the young clusters should be a good representation of the current SFR.

Using data from the literature as well as archival HST imaging, we have further tested the high SFR regime of the relation.  We found, in agreement with the results of Wilson et al.~(2006), that the relation continues to hold up to SFRs of a few hundred solar masses per year.

We have also presented a schematic model (following on Weidner et al.~2004) to explain why some dwarf (post) starburst galaxies lie on the left of the observed relation.  The basic model is that the formation of a massive cluster is enough to disturb a substantial part of the galaxy, and cause an ISM blowout (as seen in NGC~1569) which effectively terminates further star-formation.  This same process is expected to take place within star-forming complexes in spiral galaxies.

By generating a large sample of Monte Carlo simulations of cluster populations with varying underlying initial mass functions and cluster-formation rates, we have shown that the observed \relation~ relation is best fit by a Schechter mass function with \mstar$=1-5\times10^{6}~\msun$.  This rather high value of \mstar~ makes it extremely difficult to detect a difference between a pure power-law or a Schechter type mass function directly (i.e. through binned cluster mass distributions).  However, size-of-sample effects which rely heavily on the most massive or brightest object in a sample are much more sensitive to small differences between these two functions at the high mass end.  We find that the data are inconsistent with the pure power-law, $\alpha=2$ case at the $>8\sigma$ level.

Using this set of simulations, we also showed that a constant fraction of star-formation goes into clusters.  This fraction, which we term $\Gamma$, is $0.08\pm0.03$, meaning that optically selected clusters represent only 8\% of the total star-formation in a galaxy.  This is similar to that found for the solar neighborhood by Lada \& Lada~(2003) and Lamers \& Gieles~(2007),  and it is also in fair agreement with that derived for the SMC, namely $2-4$\% (Gieles \& Bastian~2008).  $\Gamma$ appears to be independent of the star-formation rate over 6 orders of magnitude.  From this we conclude that cluster formation does not have 'multiple modes' and that the cluster formation history of a galaxy should accurately reflect its star-formation history (once a proper accounting of the cluster formation rate and variance in the stellar fields is taken into account).

Using the fit to the \relation~ relation by Weidner et al.~(2004), and extrapolating to higher SFRs, we have estimated the peak SFR and infrared luminosity of seven (post-starburst) galactic mergers by estimating the absolute magnitude of the brightest star cluster when corrected for stellar evolutionary fading.  We estimate SFRs between 50 and $>4000$~\msunyr~ and infrared luminosities between $10^{11.5} - 10^{13.4}$~\lsun.  These are lower limits as we have not corrected for cluster mass-loss (fading) due to tidal fields and internal dynamics. Additionally, the infrared luminosities are also lower limits as  they do not take into consideration of any additional contribution to the \lir~ by an AGN component.  These results further demonstrate that many mergers pass through a LIRG/ULIRG/HLIRG phase during their evolution.

NGC~34, which currently has a high SFR ($>75$~\msunyr), also hosts a population of bright massive clusters with ages of $\sim150-600$~Myr (Schweizer \& Seitzer~2007).  Using these clusters we have estimated a peak SFR of $\sim600$~\msunyr~ which corresponds to an infrared luminosity of $10^{12.5}$~\lsun, well within the ULIRG regime.  Thus we have placed a lower-limit to the duration of the LIRG/ULIRG phase of 150~Myr for this galaxy.  

\section*{Acknowledgments}

NB thanks Mark Gieles for helpful discussions and for providing the stochastic power-law and Schechter function sampling programmes, used throughout this work.  Fran\c{c}ois Schweizer, Cathie Clarke, Thomas Maschberger, \& John Hibbard, are gratefully acknowledged for interesting and helpful discussions. The anonymous referee is acknowledged for helpful suggestions. This paper is based in part on observations with the NASA/ESA {\it Hubble Space Telescope}\/ which is operated by the Association of Universities for Research in Astronomy, Inc. under NASA contract NAS5-26555.

\bsp
\label{lastpage}
\end{document}